# Magnetic field generation in plasma waves driven by co-propagating intense twisted lasers

Y. Shi,[1*] J. Vieira,[2] R. M. G. M. Trines,[3] R. Bingham,[3,4] B.F. Shen,[5,6] R. J. Kingham[1]

[1] *Blackett Laboratory, Imperial College London, London SW7 2AZ, United Kingdom*
[2] *GoLP/ Instituto Superior Técnico, Universidade de Lisboa, Lisbon, Portugal*
[3] *Central Laser Facility, STFC Rutherford Appleton Laboratory, Didcot, OX11 0QX, United Kingdom*
[4] *Department of Physics, University of Strathclyde, Strathclyde, G4 0NG, United Kingdom*
[5] *State Key Laboratory of High Field Laser Physics, Shanghai Institute of Optics and Fine Mechanics, Chinese Academy of Sciences, Shanghai 201800, China*
[6] *Department of Physics, Shanghai Normal University, Shanghai 200234, China*

**Abstract**

We present a new magnetic field generation mechanism in underdense plasmas driven by the beating of two, co-propagating, Laguerre-Gaussian (LG) orbital angular momentum (OAM) laser pulses with different frequencies and also different twist indices. The resulting twisted ponderomotive force drives up an electron plasma wave with a helical rotating structure. To second order, there is a nonlinear rotating current leading to the onset of an intense, static axial magnetic field, which persists over a long time in the plasma (ps scale) after the laser pulses have passed by. The results are confirmed in three-dimensional particle-in-cell simulations and also theoretical analysis. For the case of 300 fs duration, $3.8 \times 10^{17}$ W/cm$^2$ peak laser intensity we observe magnetic field of up to 0.4 MG. This new method of magnetic field creation may find applications in charged beam collimation and microscale pinch.

Since the invention of high power lasers in the 1970's, laser created plasmas have been widely studied and developed into a broad range of applications, ranging from particle accelerators [1] and X-ray sources to inertial confinement fusion. Most of these studies and applications rely on energy and linear momentum coupling from laser to plasma. Magnetic field creation plays an important role in laser plasma interaction. The most well-known methods of laser-driven DC magnetic field generation in underdense plasma are the Inverse Faraday (IF) effect for circularly polarized light [2-5] and generation in wakefields by nonlinear effects [6,7]. For laser irradiated solid density targets, other mechanisms exist such as ponderomotive generation of giga-gauss strength surface magnetic field [8-10] and mega-gauss strength fields in the bulk due to propagation of relativistic electron beams [11,12]. These self-generated fields are beneficial to applications such as charged beam collimation and microscale pinch [4-7,13,14].

However light can also possess orbital angular momentum (OAM) [15] and thereby have the potential to create plasmas with OAM. Every photon in a circularly polarized light beam has a spin angular momentum of $\hbar$. Conversely coherent light with a helical wavefront possesses OAM which is distinct from spin angular momentum. Helical wavefronts can be represented in a basis set of orthogonal Laguerre-Gaussian (LG) modes and a photon in a LG mode with a twist index of $l$ has $l\hbar$ of OAM. Whilst generation and application of such twisted light (e.g. light tweezers [16]) is well established in conventional optics at low intensities, it has only recently started to be explored at high intensities ($I > 10^{16}$ W/cm$^2$) where the optical medium is necessarily plasma. There have been some recent studies of interactions between intense LG mode laser beams and plasma. Various new simulation phenomena and theories have been proposed [17-23]. Viera et al. have shown that plasmas can obtain OAM as a result of OAM conversion in laser interactions with under-dense plasmas [18]. Shi et al. have demonstrated that plasma can acquire high OAM density in a scheme to create relativistic intensity LG modes by reflection of a laser pulse from a foil with a fan structure, using simulations and theory [17]. Mendonca et al. predict a kind of twisted longitudinal electron plasma wave which carries OAM [24]. Their method is based on finding a solution of the electrostatic paraxial equation in terms of orthogonal LG functions. Recently, experiments have reported generation of intense OAM light reaching an intensity of $10^{19}$ W/cm$^2$ [25] and 'plasma holograms', a forked diffraction grating written onto ablating plasma by laser prepulse with OAM [26].

Here we propose, for the first time, generation of an electron plasma wave with a helical rotating structure, shown in Fig. 1(c), which is driven by co-propagating, beating, LG laser

pulses – which possess OAM – (see Fig. 1(a)) and can create a static, axial magnetic field, as depicted in Fig. 1(b). This new mechanism is seen in particle-in cell (PIC) simulations and explained using a theoretical model involving electron high order fluid equations, which yields the higher order corrections responsible for B-field generation. An outline of the mechanism is as follows. The beating OAM lasers exert a ponderomotive force with a twisted profile on the electrons. After the laser pulses have passed by, electrons are left oscillating on elliptical orbits in the transverse plane with an azimuthally dependent phase offset. This collectively yields a persistent, rotating wave structure. Associated with this is a nonlinear electrical current, essentially in the configuration of a solenoid that creates a magnetic field.

This mechanism bears some similarity to quasi-static magnetic field generation in laser-driven plasma waves by nonlinear currents [6,7] but, importantly, involves OAM. Studies of inverse Faraday (IF) magnetic field creation have concentrated on the use of circular polarized laser beams with a Gaussian mode (e.g. [3-5]) though there is some limited amount of theoretical work on linearly polarized beams with a LG mode [2]. The effectiveness of IF depends strongly on the laser absorption coefficient (into the plasma) which is usually very low (and arguably not well understood) so that IF usually requires a relatively long time to create strong fields. In our work, the details of theory and the simulation results will demonstrate a different method of magnetic field creation, outlined above, which can persist a long time (ps scale for the conditions here) after the laser interaction ceases. The longer term decay of this magnetic field can be followed based on the knowledge of plasma wave evolution, in principle. The new magnetic field generation mechanism presented here will benefit the same applications that use IF fields, such as improving the quality of laser-produced electron beams [4].

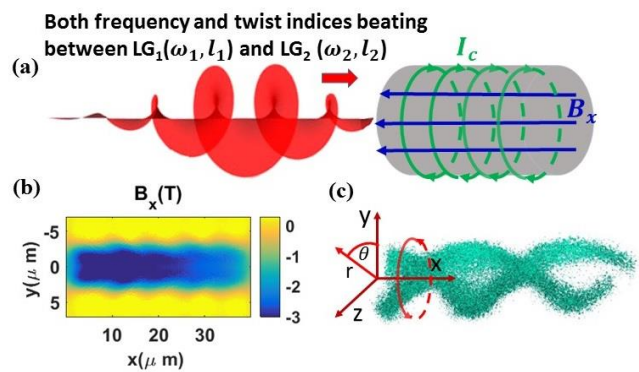

Fig. 1 (colour online). (a) Illustration of the scheme to generate plasma waves with helical rotating structure by two co-propagating OAM lasers beating both in frequency and twist index.

(b) The axial magnetic field $B_x$ in x-y plane (at z = 0). (c) The double helical electron density $n_e$ rotating around x-axis in 3D. The time is 600 fs after the laser ($\tau$ =160 fs) has passed by.

To obtain the results presented in Fig. 1, we performed three-dimensional particle-in cell (PIC) simulations using EPOCH [21] with the following parameters. The frequencies and twist indices of the two laser beams are $\omega_1 = \omega_0$, $\omega_2 = 0.95\omega_0$ and $l_1$ = -1 , $l_2$ = +1, respectively, so that the frequency difference is the same as the plasma frequency $\omega_p = 0.05\omega_0$. Here $\omega_0$ is the frequency corresponding to 800nm wavelength and $\omega_p = \sqrt{4\pi n_0 e^2/m_e}$ is the plasma frequency, where $n_0$ is the electron density, $e$ is the electron charge and $m_e$ is the electron mass. A fully ionized hydrogen plasma is used with uniform density $n_0 = 4.5 \times 10^{18}$ cm$^{-3}$ and zero temperature, initially. Both laser pulses are Gaussian time-enveloped, linearly polarized along the y-axis and propagate along the positive x-direction. Both have a diameter of $w$ = 5μm ($1/e^2$ intensity measure), pulse duration of $\tau_g$ = 160fs and peak amplitude of $E_p = 4.8 \times 10^{11}$ V/m (peak vacuum intensity is $I_p$ = 3.1×10$^{16}$ W/cm$^2$). This intensity corresponds to a peak dimensionless vector potential of $a_0 = 0.2$ where $a_0 = eE_0/m_e\omega_0 c$. The temporal pulse envelope is truncated once the intensity drops by $1/e^2$. The simulation box is 40μm×25μm×25μm in the $x \times y \times z$ directions, respectively. The simulation mesh size is d$x$ = d$y$ = d$z$ = 0.05 μm. The total number of macro particles per cell is 4. A schematic of the system and some key simulation results are given in Fig.1. Of particular note are a static axial magnetic field $B_x$ in x-y plane, shown in Fig. 1(b), and the double helical electron density distribution $n_e$ presented in Fig. 1(c). As time moves on, the double helical density rotates around the x-axis with angular frequency $\omega_p/2$. The scheme works for other choices of twist indices. In particular, frequency beating between a LG and a Gaussian beam is viable and is likely to by easier to realize experimentally, but leads to a more complex theoretical analysis. Therefore, opposite twist indices are chosen in the following analysis to elucidate the physics and verify the PIC results. We have performed PIC simulations changing box size, laser pulse, laser spot and boundary conditions. All results show the phenomena in our simulations are robust.

Firstly, we solve the non-relativistic cold electron-fluid equations with beating, LG EM waves as the driving source. (Gaussian units are used.) In LFT [27], the linearized electron fluid momentum, mass–continuity and Poisson's equations describing a laser-driven electron plasma wave can be written, respectively, as

$$\begin{cases} m_e \frac{\partial \vec{u}}{\partial t} = e\nabla\phi + \vec{f_L} \\ \frac{\partial \delta n_e}{\partial t} + n_0 \nabla \cdot \vec{u} = 0 \\ \nabla^2 \phi = 4\pi e \delta n_e \end{cases} \quad (1)$$

where $n_0$ here is the number density of ions (charge Z=1) that are assumed to be immobile and uniformly distributed in space, $\delta n_e$ is the difference between the ion and electron densities, $\vec{u}$ is the velocity of the electron fluid, $\phi$ is the electrostatic potential and $\vec{f_L}$ is the ponderomotive force of the lasers. A cold electron fluid is assumed. LFT is valid for small responses; $\{e\phi/(m_e c^2), u/c, \delta n_e/n_0\} \ll 1$. The electric field $E_y(r, \theta, x, t)$ of each LG mode is $E_{1,2} = C_l E_0 (\sqrt{2}r/w)^{|l|} exp(-r^2/w^2) cos(\omega_{1,2} t - k_{1,2} x \pm l\theta)$ where $\omega_p = \omega_1 - \omega_2, k_p = k_1 - k_2$, and the subscripts 1 and 2 label each pulse. Here, $C_l$ is a normalisation constant [15]. Also, $r$, $\theta$, $x$ denote cylindrical polar coordinates about the $x$-axis, as depicted in Fig. 1(c). The slowly varying envelop of each pulse is contained in $E_0(x - v_g t)$ and we assume equal group velocities for each mode. The resulting ponderomotive force from the superposition of the (synchronised) pulses is $\vec{f_L}(r, \theta, x, t) = -[C_l^2 2^{|l|} e^2 E_0^2/(m_e \omega_0^2)] \nabla[N(r)(1 + cos\Phi)/2]$, where $\Phi = \omega_p t - k_p x + 2l\theta$ and $N(r) = exp(-2r^2/w^2)(r/w)^{2|l|}$. The ponderomotive force $\vec{f_L}$ has an azimuthal component, in contrast to the case of beating, Gaussian modes (which have $l=0$). When the laser pulse duration $\tau$ is much larger than the plasma period $T_p$, the solutions of electron density $\delta n_e(r, \theta, x, t)$ after the laser interaction can be approximated as

$$\delta n_e/n_0 = -0.5\, \eta N(r) Y(r) sin\Phi \text{ with } Y(r) = 8r^2/w^2 - (k_p w)^2/2 - (4 + 8|l|), \quad (2)$$

where the key, dimensionless parameter is $\eta = C_l^2 2^{|l|} a_0^2 (\omega_p \tau)/(k_p w)^2$. Here $\eta \propto a_0^2 \tau/w^2$ decides the amplitude of plasma wave and particle oscillation, and is used as an expansion parameter for a subsequent higher order analysis. In the PIC simulation shown in Fig. 1, $\eta \sim 0.3$. $N(r)$ and $Y(r)$ are dimensionless radial shape functions. The electrostatic field components $E_x, E_r, E_\theta$ and velocity of the electric fluid $u_x, u_r, u_\theta$ can also readily be calculated. They are also linear in $\eta$ and $N(r)$. The important result that will be used later to determine the current supporting the magnetic field is

$$u_\theta(r, \theta, x, t) = u_{\theta 1} sin\Phi \text{ where } u_{\theta 1} = 0.5\eta \omega_p w^2 N(r) l/r. \quad (3)$$

Other formulae are given in Supplemental Material [28]. For simplicity, Eq. (2) and (3) have been obtained assuming a flat-top pulse with duration $\tau$. For the truncated Gaussian pulse used in the PIC simulation (i.e. $I(t) = 0$ for $|t - t_p| \geq \tau_g/2$ where $t_p = \tau_g/2$ is the time of peak intensity), $\tau = 0.75\tau_g$ is appropriate. We have verified that the plasma wave growth and B-field generation are not significantly influenced by the abrupt truncation of the temporal profile. A simulation where the Gaussian pulse is truncated three times further out (i.e. $3\tau_g/2$ from the centre), shows a slightly larger $B_x$, consistent with the increased pulse energy. The length of the twisted plasma wave would reasonably be expected to be on the order of the Rayleigh length, which is longer than our simulation window. So diffraction of the laser beams is not considered here.

PIC simulation and theory results of electron density $\delta n_e/n_0$, time averaged rotating current $\langle j_\theta \rangle$ and axial magnetic field $B_x$ are shown in Fig. 2 and Fig. 3, respectively, 600 fs after the laser pulse has left the simulation box. The LFT prediction for $\delta n_e$ (and also $\vec{u}$ and $\vec{E}$ which are not shown here) after the laser has passed by agrees well with the PIC simulation results, both qualitatively and quantitatively, as seen by comparing Fig. 2(a) and Fig. 3(a). In the LFT the electric current density comes from $\vec{j^e} = -en_0\vec{u}$, and the displacement current density is equal to $\vec{j^{dis}} = (\partial \vec{E}/\partial t)/4\pi = en_0\vec{u}$. The net current used to calculate magnetic field should be the sum of these currents, which vanishes in the LFT (see Supplemental Material [28] for more details). The simulation results, on the other hand, show a time-averaged net current in the azimuthal direction $\langle j_\theta \rangle = \langle j_\theta^e + j_\theta^{dis} \rangle$ (i.e. net rotating current), which is almost $10^{12}$ A/m$^2$ (see Fig. 2 (b) ). This equates to a normalised current of $\langle j_\theta \rangle/(en_0c) \sim 0.004$ and indicates that it is a higher order effect, since a conduction current amplitude of $j_\theta^e/(en_0c) \sim 0.07$, is expected from LFT. A simple calculation shows that the background electrons need to rotate at a velocity of $10^6$ m/s to get such a strong rotating current density, which is rare compared with strong linearly current density. The corresponding angular velocity is $5 \times 10^{11}$ rad/s, which is 3 orders of magnitude higher than the equivalent angular velocity in the 'Light Fan' [17]. The area-averaged axial magnetic field is almost 3T (i.e. 30 kG) in our simulation.

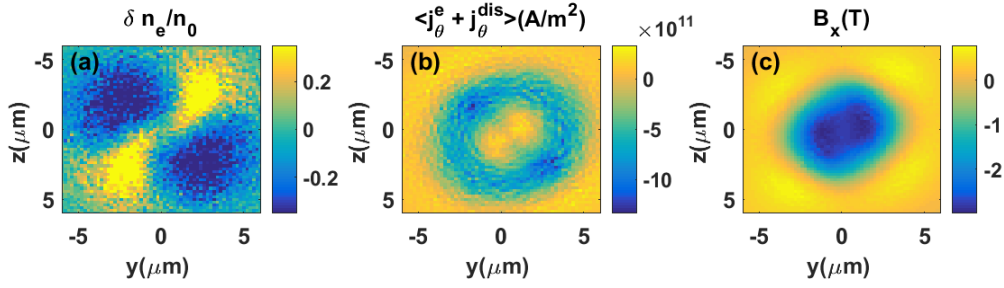

Fig. 2 (colour online). PIC results of transverse profile of (a) $\delta n_e$, (b) time averaged, azimuthal component of the net current density $\langle j_\theta \rangle = \langle j_\theta^e + j_\theta^{dis} \rangle$, and (c) $B_x$, all at $x = 20$ µm and 600fs after the laser has passed by.

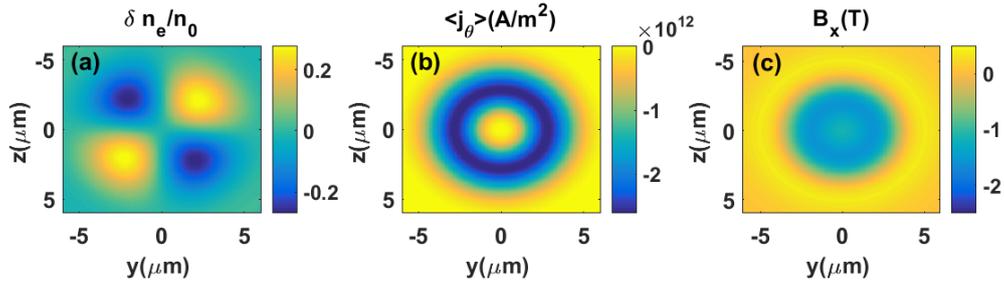

Fig. 3 (colour online). Linear fluid theory (LFT) predictions, for the same situation considered in Fig. 2. (a) $\delta n_e$, (b) higher order, time-averaged $\langle j_\theta \rangle$, and (c) $B_x$ predicted from the solenoid model (i.e. from $\langle j_\theta \rangle$).

The result of the LFT is a plasma wave with a helical rotating structure and which carries OAM. It is very different from the longitudinal plasma wave driven by the beating of two Gaussian lasers, where transverse profiles depend only on radius. Additionally, it is not the same as a longitudinal plasma wave with a twisted phase front [24], as proposed by Mendonca et al. There, a solution of the electrostatic paraxial equation in terms of LG functions is given. In our case, Eq. (2), describing the wave's density perturbation, does not involve use of the paraxial approximation. The primary focus of this paper is axial magnetic field creation, which according to our PIC simulations are associated with a net rotating current. Since the linear fluid theory predicts zero net azimuthal current, a higher order calculation is required to describe the generation of axial magnetic field by the helical plasma wave. According to L.M. Gorbunov's higher order fluid theory[6,13], the second order (in parameter $\eta$) current exists and can be calculated from electron density and velocity in LFT. See the Supplemental Material [28] for more information. With these results we obtain the net current density needed to find the magnetic field $j_\theta = -en_1 u_{\theta 1}\sin^2\Phi$, where $n_1$ and $u_{\theta 1}$ are the amplitudes of the first-order electron density and velocity perturbations from LFT. Averaged over time $t$ (i.e. the plasma period) and angle $\theta$ (i.e. over $2\pi$) we find

$$\langle j_\theta \rangle = en_0 c l\, K\, (C_l a_0)^4 \left(\frac{c\tau}{w}\right)^2 \frac{4^{|l|}}{k_p w}, \tag{4}$$

which scales as $\langle j_\theta \rangle/(en_0 c) \propto \eta^2(k_p w)$, where $K = (|l| + 1/2 - \epsilon)e^{-4\epsilon}\epsilon^{2|l|-1/2}$ and $\epsilon = (r/w)^2$. This prediction for $\langle j_\theta \rangle$ is shown in Fig. 3(b) as rotating current $\langle j_\theta \rangle$, and is close to the simulation result $\langle j_\theta^e + j_\theta^{dis} \rangle$ in Fig. 2(b). An important point is that this theory links $\langle j_\theta \rangle$ to the twist index $l$ of the laser beam drive. In particular notice that $\langle j_\theta \rangle = 0$ when $l = 0$. This demonstrates that OAM is essential to the magnetic field creation in our scheme.

Now we calculate the magnetic field from the time-averaged net rotating current. To simplify the analysis, we restrict attention to motion in the transverse plane which is valid in the limit $u_x \ll u_\perp, E_x \ll E_\perp$ which requires $k_p r/2 \ll |l|$. More information on the approximation can be found in Supplemental Material [28]. This approximation can always be met by choosing sufficiently low density plasma or a narrow laser beam. The PIC simulations satisfy these conditions. Given that $\langle j_\theta \rangle$ represents a solenoidal current which is distributed in $r$, and ignoring end effects, $B_x(r) = (4\pi/c) \int_r^\infty \langle j_\theta \rangle dr'$. This theoretical result for $B_x$, presented in Fig. 3(c), compares well with the simulation result shown in Fig. 2(c). To obtain a scaling relation between the axial magnetic field and the laser parameters, we use an effective solenoidal current per unit length of $I_{sol} = \int_{r_1}^{r_2} \langle j_\theta \rangle dr$ in the equation for the B-field inside an ideal solenoid, $B_{sol} = (4\pi/c)I_{sol}$, and choose $r_1 = 0.4w$ and $r_2 = 0.7w$ to capture the bulk contribution of the ring-like current density profile. We find that the magnetic field inside the solenoid scales as $B_{x,[T]} \sim -27\tau_{[ps]}^2 I_{[10^{16}]}^2$ for the twist indices, plasma density and beam radius used in the PIC simulations, where $I_{[10^{16}]}^2$ denotes the peak laser intensity ($I_p$) in units of $10^{16}$ W/cm$^2$ and the B-field and pulse duration are in units of Tesla and ps, respectively. More generally, assuming that the distributed solenoid current profile scales with beam radius, $\sim w \langle j_\theta \rangle_{\text{peak}}$, the magnetic field is expected to scale as $B \sim n_0^{1/2} \tau^2 I_p^2/w^2$.

To compare with the simulation, the $B_x(y,z)$ profile from the PIC calculation, half way along the structure at $x = 20$ μm (see Fig. 1(b)) is averaged within $r = 1$ μm to obtain $\langle B_x \rangle_{PIC}$ for a range of simulations where either pulse intensity or pulse length is varied from the default configuration (used for Fig. 2). Fig 4(a) shows that the theoretical scaling result (when calibrated by a factor of ¾ ) agrees well with the 3D PIC calculation for peak intensities below $I_p = 1.9 \times 10^{17} \text{W/cm}^2$, with the PIC result following the $\langle B_x \rangle \propto I_p^2$ scaling in this limit. The PIC results also display the predicted $\langle B_x \rangle \propto \tau^2$ scaling for short pulses below 200 fs and compare well quantitatively with the calibrated model in this limit. As $\tau^2 I_p^2$ gets bigger, the axial magnetic field from the PIC simulations increases more slowly than the value predicted by the theoretical scaling relation and approaches saturation. This discrepancy is likely due to nonlinear effects that become important. For instance, $\delta n_e/n_0$ already exceeds 0.2 in the default simulation shown in Fig. 2. Going beyond this, third order (in $\eta$) and higher corrections would become significant. We note that the theoretical model also predicts $I_c \propto \sqrt{n_0}$ which implies a stronger magnetic field for higher plasma density.

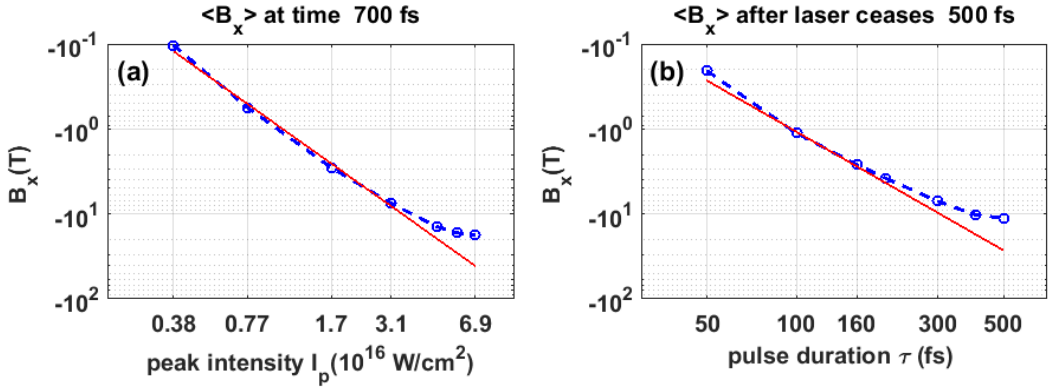

Fig. 4 (colour online). Comparison of <$B_x$> found in 3D PIC simulations (blue circles connected by dashed lines) with the theory prediction $B_x$ (T) ~ $-20\tau^2_{[ps]}I^2_{[10^{16}]}$ (red lines) using a logarithmic scale. (a) Varying $I_p$ (laser peak intensity) at fixed pulse duration ($\tau$ = 300fs). (b) Varying $\tau$ (laser pulse duration) at fixed $I_p$ = 3.1×10$^{16}$ W/cm$^2$ and measured promptly after the laser drive has ceased. Here, <$B_x$> is averaged over 0≤ $r$ ≤ 1 μm at $x$ = 20 μm.

The OAM laser-plasma magnetic field generation scheme presented here has benefits over the inverse Faraday mechanism, particularly in the sub-picosecond duration regime. We have carried out a comparable 3D PIC simulation of IF under identical conditions to those used in Fig. 1 and 2, except that a single circular polarised Gaussian laser with the same power (as the combined beams in our scheme) was used. The axial magnetic field after the laser has passed by is almost 2 orders of magnitude lower than produced by the OAM-beat mechanism in Fig.

1(b). This demonstrates that the OAM-beating mechanism reported in this paper can be more efficient than IF for short laser pulses. Finally, the scheme can also be seen as creation of a laser created $\theta$-pinch, on the micrometre and nanosecond scale. Others have reported creation of a nanometer scale Z-pinch from laser irradiated nanowire arrays [14].

In conclusion, a new method of axial magnetic-field generation in underdense plasma, based on intense laser pulses with orbital angular momentum (OAM), is proposed. It utilizes two co-propagating, sub picosecond duration, laser pulses with different Laguerre-Gauss (LG) modes which beat in both frequency and twist index *l*. This work considers pulses below the relativistic intensity limit. Three-dimensional PIC simulations and supporting theoretical analysis show that the beating OAM laser beams create a twisted ponderomotive force that transfers OAM to the driven electron wave structure. At the same time, Viera et al. have shown that a light spring with a helical spatiotemporal intensity profile can bring OAM and new topological control to Laser-Plasma accelerators [30]. The plasma wave in our paper has some special characteristics such as a helical, rotating electron density and also a net nonlinear solenoidal current that creates the axial magnetic field. It is different from a 3D longitudinal plasma wave driven by the beat of laser beams with a transverse Gaussian mode. The theory presented here is based on linear fluid equations combined with a higher order fluid theory. It reveals that the 2$^{nd}$ order current exits and also that the key parameters for this current, and hence B-field, are *l* and $\tau^2 I_p^2$. This establishes that OAM is essential to this magnetic field source. Compared to other laser based methods of magnetic field creation in underdense plasma such as the inverse Faraday effect, the magnetic field in our scheme can be much stronger and persist a long time, even after the laser interaction stops. Higher magnetic fields should be possible if multiple laser beams are used or higher density plasma is considered. Such a static magnetic field may find some applications to charged beam collimation or microscale pinch. Our simulations show that this OAM-laser based magnetic field source is also possible via the beating between a LG and a Gaussian beam, which may be easier for a proof-of-principle experiment.

Dr. Y. Shi is a Newton International Fellow. This work is supported by the Royal Society. This work used the ARCHER UK National Supercomputing Service (http://www.archer.ac.uk) and the Imperial College Research Computing Service (https://doi.org/10.14469/hpc/2232 ). J. V. acknowledges the support of FCT (Portugal) Grant No. SFRH/IF/01635/2015, and B.F.S the support from the Ministry of Science and Technology of the People's Republic of China (Grant No.2016YFA0401102, 2018YFA0404803) and the Strategic Priority Research Program of the Chinese Academy of Sciences (Grant No. XDB16).

* To whom all correspondence should be addressed: yin.shi@imperial.ac.uk**Reference**

[1] E. Esarey, C. B. Schroeder, and W. P. Leemans, Rev. Mod. Phys. 81, 1229 (2009).
[2] S. Ali, J. R. Davies, and J. T. Mendonca, Phys. Rev. Lett. **105**, 035001 (2010).
[3] M. G. Haines, Phys. Rev. Lett. **87**, 135005 (2001).
[4] Z. Najmudin, M. Tatarakis, A. Pukhov, E. L. Clark, R. J. Clarke, A. E. Dangor, J. Faure, V. Malka, D. Neely, M. I. K. Santala and K. Krushelnick, Phys. Rev. Lett. **87**, 215004 (2001).
[5] Z. M. Sheng and J. Meyer-ter-Vehn, Phys. Rev. E **54**, 1833 (1996).
[6] L. Gorbunov, P. Mora, and T. M. Antonsen, Phys. Rev. Lett. **76**, 2495 (1996).
[7] Z. M. Sheng, J. Meyer-ter-Vehn, and A. Pukhov, Phys. Plasmas **5**, 3764 (1998).
[8] S. C. Wilks, W. L. Kruer, M. Tabak, and A. B. Langdon, Phys. Rev. Lett. **69**, 1383 (1992).
[9] R. N. Sudan, Phys. Rev. Lett. **70**, 3075 (1993).
[10] M. Tatarakis, I. Watts, F. N. Beg, E. L. Clark, A. E. Dangor, A. Gopal, M. G. Haines, P. A. Norreys, U. Wagner, M.-S. Wei, M. Zepf and K. Krushelnick, Nature **415**, 280 (2002).
[11] A. P. L. Robinson, D. J. Strozzi, J. R. Davies, L. Gremillet, J. J. Honrubia, T. Johzaki, R. J. Kingham, M. Sherlock, and A. A. Solodov, Nuclear Fusion **54**, 054003 (2014).
[12] J. R. Davies, A. R. Bell, and M. Tatarakis, Phys. Rev. E **59**, 6032 (1999).
[13] L. M. Gorbunov, P. Mora, and T. M. Antonsen, Phys. Plasmas **4**, 4358 (1997).
[14] V. Kaymak, A. Pukhov, V. N. Shlyaptsev, and J. J. Rocca, Phys. Rev. Lett. **117**, 035004 (2016).
[15] L. Allen, M. W. Beijersbergen, R. J. C. Spreeuw, and J. P. Woerdman, Phys. Rev. A **45**, 8185 (1992).
[16] A. M. Yao and M. J. Padgett, Adv. Opt. Photonics **3**, 161 (2011).
[17] Y. Shi, B. Shen, L. Zhang, X. Zhang, W. Wang, and Z. Xu, Phys. Rev. Lett. **112**, 235001 (2014).
[18] J. Vieira, R. M. G. M. Trines, E. P. Alves, R. A. Fonseca, J. T. Mendonça, R. Bingham, P. Norreys, and L. O. Silva, Phys. Rev. Lett. **117**, 265001 (2016).
[19] L. Zhang, B. Shen, X. Zhang, S. Huang, Y. Shi, C. Liu, W. Wang, J. Xu, Z. Pei and Z. Xu, Phys. Rev. Lett. **117**, 113904 (2016).
[20] X. Zhang, B. Shen, Y. Shi, X. Wang, L. Zhang, W. Wang, J. Xu, L. Yi, and Z. Xu, Phys. Rev. Lett. **114**, 173901 (2015).
[21] J. Vieira and J. T. Mendonça, Phys. Rev. Lett. **112**, 215001 (2014).
[22] W. Wang, B. Shen, X. Zhang, L. Zhang, Y. Shi, and Z. Xu, Sci. Rep. **5,** 8274 (2015).
[23] X. Zhang, B. Shen, L. Zhang, J. Xu, X. Wang, W. Wang, L. Yi, and Y. Shi, New J. Phys. **16**, 123051 (2014).
[24] J. T. Mendonca, S. Ali, and B. Thidé, Phys. Plasmas **16**, 112103 (2009).
[25] A. Denoeud, L. Chopineau, A. Leblanc, and F. Quéré, Phys. Rev. Lett. **118**, 033902 (2017).
[26] A. Leblanc, A. Denoeud, L. Chopineau, G. Mennerat, P. Martin, and F. Quere, Nat. Phys. **13**, 440 (2017).
[27] R. Fedele, U. de Angelis, and T. Katsouleas, Phys. Rev. A **33**, 4412 (1986).
[28] See Supplemental Material [url] for details on the LFT theory and the higher order fluid theory, which includes Ref. [29].
[29] J. M. Dawson, Physical Review **113**, 383 (1959).
[30] J. Vieira, J. T. Mendonça, and F. Quéré, Phys. Rev. Lett. **121**, 054801 (2018).11